# Observation of collective atomic recoil motion in a momentum-squeezed, ultra-cold, degenerate fermion gas

Pengjun Wang[1], L. Deng[2], E.W. Hagley[2], Zhengkun Fu[1], and Shijie Chai[1], Jing Zhang[1]*

[1] State Key Laboratory of Quantum Optics and Quantum Optics Devices, Institute of Opto-Electronics, Shanxi University, Taiyuan 030006, China

[2] Physics Laboratory, National Institute of Standards & Technology, Maryland USA 20899

* Email: jzhang74@sxu.edu.cn; jzhang74@yahoo.com

Date: June 14, 2010

## Abstract

We demonstrate clear collective atomic recoil motion in a dilute, momentum-squeezed, ultra-cold degenerate fermion gas by circumventing the effects of Pauli blocking. Although gain from bosonic stimulation is necessarily absent because the quantum gas obeys Fermi-Dirac statistics, collective atomic recoil motion from the underlying wave-mixing process is clearly visible. With a single pump pulse of the proper polarization, we observe two mutually-perpendicular wave-mixing processes occurring simultaneously. Our experiments also indicate that the red-blue pump detuning asymmetry observed with Bose-Einstein condensates does not occur with fermions.

Collective atomic recoil motion modes produced by coherent electromagnetic waves can only be directly observed in an ensemble of ultra-cold atoms where the recoil momentum is greater than the intrinsic momentum spread of the atomic cloud. Such collective recoil modes were first observed by illuminating the short axis of a Bose-Einstein Condensate (BEC) with a red-detuned long duration laser pulse having the proper polarization. The resulting scattering pattern of atoms was distinctive, unidirectional and highly regular [1]. Since that seminal experiment it has been argued, and also widely accepted, that the underlying physics of this intriguing light-ultra-cold-matter interaction process is Rayleigh scattering of pump photons by a matter-wave grating produced by a spontaneous scattering process that is further amplified by bosonic stimulation (matter-wave amplification) [1-4].

Recently, Deng et al. reported the first small-signal electromagnetic wave-mixing and propagation theory [5,6], as well as a supporting experimental study [7]. These works show that matter-wave superradiance is fundamentally based on multi-matter-optical wave mixing and propagation, with bosonic stimulation providing a second gain mechanism. Although bosonic stimulation in BECs may be necessary for high gain and higher-order scatterings, it is not critical in the initial stages of this intriguing wave-mixing/propagation process. Therefore the underlying wave-mixing process should be able to be observed with a cold fermion gas. In contrast to BECs, where the spatial density variation results in an astonishing pump laser detuning sign asymmetry [7], the wave-mixing process with fermions should not depend on the sign of the laser detuning because the vapor density is uniform.

In this Letter we report the first clear experimental evidence of collective atomic recoil motion in a momentum-squeezed, degenerate fermion gas, where the mechanism of bosonic stimulation is obviously absent. We demonstrate the generation of collective atomic recoil motion when a single, unidirectional laser pulse interacts with a degenerate fermionic system, and we demonstrate the coexistence of multiple mutually perpendicular propagating matter-optical wave-mixing processes. To the best of our knowledge this is the first collective atomic recoil motion experiment using a momentum-squeezed, degenerate fermionic system.

Our experimental setup is present in [8]. We first prepare a degenerate fermion gas by cooling a mixture of $^{87}$Rb + $^{40}$K atoms using well-developed evaporative and sympathetic cooling techniques in a quadrupole-Ioffe configuration (QUIC) trap. At the end of this cooling process the $^{40}$K atomic cloud reaches Fermi degeneracy at T~ 400 nK (~ 0.3 $T_F$) with about $2 \times 10^6$ atoms. The atomic sample is detected by the time of flight absorption image (TOF) and the probe beam used for absorption imaging passes through the glass cell in the *x*-direction (gravity axis). All $^{87}$Rb atoms are then removed, resulting in approximately $2 \times 10^6$ $^{40}$K atoms in a cloud ($4S_{1/2}$, |F=9/2,$m_F$=9/2> hyperfine Zeeman state) approximately 90 μm in length and 20 μm (which is measured with 1.2 ms short TOF) in diameter [8]. The final temperature of the degenerate fermion gas at the end of this sympathetic cooling process is, however, still too high for demonstrating collective atomic recoil motion (Fig. 1a). In order to overcome this limitation we subject the atomic cloud to a controlled adiabatic expansion. This critically important step has two consequences. First, it reduces the temperature of the gas to the point where the recoil momentum is dominant. Second, and more importantly, it creates a momentum-squeezed, degenerate fermion cloud that facilitates the observation of collective atomic recoil motion for a system that must obey the Pauli exclusion principle. Specifically, we adiabatically relax the magnetic trap in a prescribed way with the asymmetrically spatial aspect ratio of the fermion cloud. Due to the significant temperature reduction accompanying adiabatic expansion, the momentum spread and distribution evolve such that they mirror the spatial distribution of the cloud. This generates an elongated, degenerate fermion cloud that expands slowly with a nearly fixed aspect ratio after being released from the magnetic trap (see Fig. 1b). At the end of adiabatic expansion the fermion cloud is 85 μm by 30 μm (after 1.2 ms TOF), and its momentum distribution is frozen with the momentum spreads along the *x*- and *y*-directions much smaller than the spread in the *z*-direction, thereby producing an ultra-cold, momentum-squeezed, degenerate fermion gas [9,10].

Due to Pauli blocking no atoms can be scattered into an already occupied quantum state where both atomic internal and external (momentum) degrees of freedom are the same.

However, Pauli blocking does not prevent fermions from occupying states with the same projection of momentum along any particular axis, as long as at least one of the other two projections is different. For instance, one can scatter an unlimited number of atoms with the same internal atomic state into the same ($k_x$, $k_y$) momentum state as long as these atoms all have different $k_z$ components to distinguish them from one another. This is precisely the momentum-squeezed state that we produce with the degenerate fermion gas. In essence, we create a collective atomic motion mode in which all scattered fermions have nearly identical momentum components in the direction of pump laser while having unique momentum components perpendicular to this direction. This results in a correspondingly longer atomic coherence length of the gas along the propagation direction of the pump laser.

Physically, the process that leads to the internally-generated electromagnetic field is based on multi-matter-optical wave mixing, where the usual nonlinear optical wave-mixing and propagation process must be generalized to include atomic recoil motion. Inside the medium excited by the pump laser, photons are first scattered by the usual spontaneous emission process. In the case of a BEC, the coherence time is long and the field propagating along the long dimension of the condensate will experience the highest gain. On the other hand, in a momentum-squeezed, degenerate fermion gas the momentum spread along the *z*-direction (long axis) is larger than along the other directions, and the correspondingly smaller coherence time along this axis results in inefficient coherent build up of the electromagnetic field generated internally by the inelastic light scattering process. Consequently, wave mixing and generation is more favorable in the *x-y* plane where the momentum spread is smaller and the coherence time is longer. The wave-mixing process occurs via a two-photon resonance transition involving the absorption of a pump photon and the subsequent spontaneous and/or stimulated emission of a photon. If the emitted photon counter-propagates the pump laser, the atom will acquire $2\hbar k$ net momentum transfer and move in the propagation direction of the pump laser (the forward-scattered component).

In Fig. 2 shows a TOF image of collective recoil motion mode after a pump laser pulse interacts with the fermion cloud. For this image the laser polarization is perpendicular to the long axis (*x*-direction) of the atomic cloud (as in the case of all BEC superradiance experiments reported [1,2,7]), and we observe the familiar pair of scattered components moving at $\pm 45^o$ (i.e., *y*$\pm$*z*) with respect to the pump laser propagation direction (Fig. 2). We note that in addition to this pair of first-order scattered components there is also a forward scattered component. It is unlikely that this forward scattered component is due to a sequential second-order process of $\pm 45^o$ scatterings, because the atomic density and the intensity of the generated light in the $\pm z$ directions are not sufficient to produce any second-order effects. Thus, we conclude that the forward component is due to a similar matter-optical-mixing process occurring at the same time along the short axis. Indeed, with the laser polarization along the *x*-direction, emission of photons in entire *y-z* plane is allowed. Photons emitted in the $\pm z$-directions contribute to the $\pm 45^o$ scattering components, and photons emitted in +*y*-direction contribute to the forward component. In essence we have multiple mutually perpendicularly-propagating, wave-mixing processes occurring simultaneously. Thus when the polarization of pump laser is parellel to the long axis (*x*-direction), only the forward component (+*y*-direction) appears.

We also investigated the pump laser detuning effect in this fermionic system. It has been shown experimentally [7] with a Bose condensate that collective atomic recoil motion modes are strongly suppressed when the laser is detuned to the high energy side (blue detuned) of the one-photon transition. With condensates, it has been argued that the spatially-varying atomic density, the mean field, ultra-slow propagation of the generated light, and an induced optical dipole force all contribute to the strong suppression of the collective recoil mode when the laser is blue detuned [7]. In a degenerate fermion gas, however, the density gradient is absent and we expect that collective atomic recoil motion modes due to the coherent wave-mixing process should be observable with both red- and blue-detuned excitation. Fig. 3 shows two sets of TOF images similar to Fig. 2b, but with blue (Fig. 3a) and red (Fig. 3b) detuned pump lasers for different pulse durations. As the images show, there is practically no difference between these two cases, as expected from the multi-matter-optical wave mixing theory when a constant density profile is assumed.

Finally, we note that because of the low atomic density the experimental conditions are in the thin-medium limit in wave-propagation theory. Thus, coherent build-up of the generated field is necessarily weak and this results in significant spontaneous emission because the internally-generated optical fields travel near the speed of light in vacuum [6]. Never-the-less, the distinctive scattered components in (*y*$\pm$*z*) and in +*y* directions generated by the underlying wave-mixing process are unmistakable.

In conclusion, we have demonstrated clear collective atomic recoil motion in a momentum-squeezed, degenerate fermion gas by circumventing the effects of Pauli blocking for a quantum gas that obeys Fermi-Dirac statistics. We believe that the experimental studies reported here clearly establish that (1) quantum statistics are not critical in the generation of collective modes [9,11,12,13], (2) matter-optical wave mixing is the underlying physical mechanism in this single-pulse, inelastic light scattering process, and (3) there is no red-blue pump laser detuning asymmetry with a fermionic gas having a uniform density profile.

**Acknowledgement** Jing Zhang wishes to thank the National Natural Science Foundation of China (Grant No. 10725416, 60821004) National Basic Research Program of China (Grant No. 2006CB921101) for financial support. L. Deng thanks the Center for Cold Atom Physics, the Wuhan Institute of Physics and Mathematics, and the Chinese Academy of Sciences for support.

**Figures and captions**

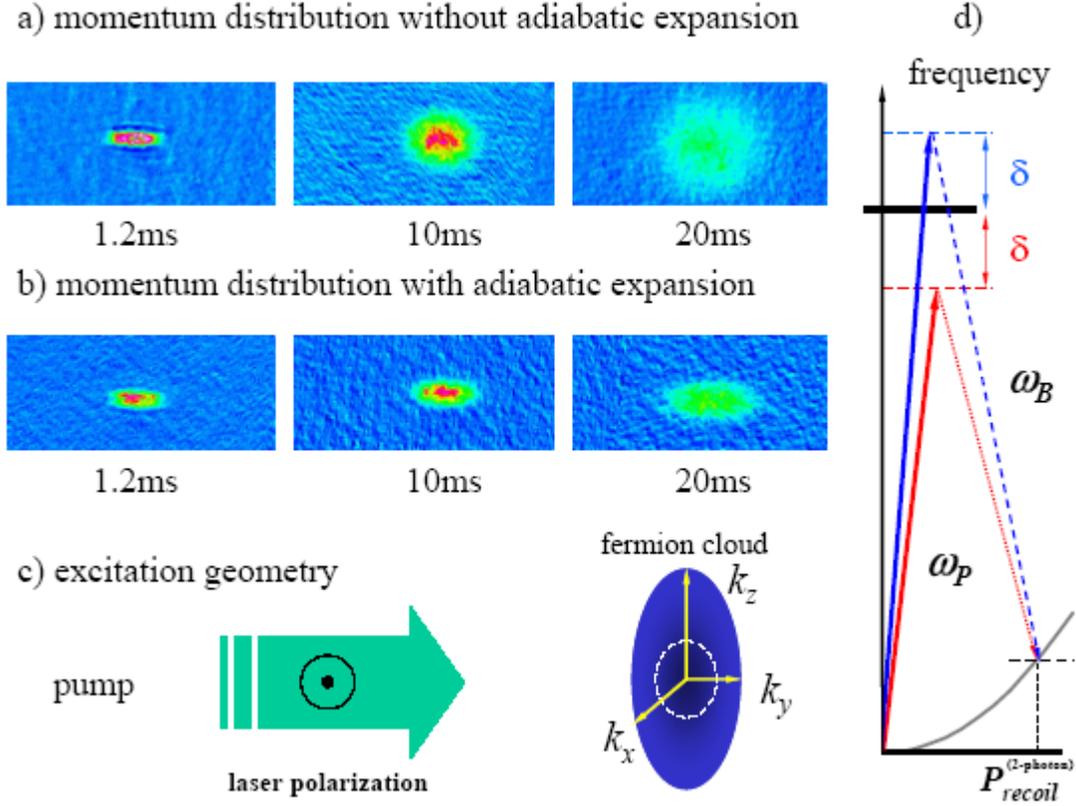

**Figure 1.** Generation of a momentum squeezed, degenerate fermion gas. a) Rapid free expansion of a released fermion cloud that did not undergo a controlled adiabatic expansion. The time in each image is the free expansion time, i.e., the time between releasing atoms from the trap and imaging. Note that after 20 ms TOF the fermion cloud is almost spherical. b) Free expansion of a fermion cloud after a controlled adiabatic expansion. Here the much colder degenerate fermion cloud expands with an almost frozen aspect ratio and momentum distribution. After 20 ms TOF, the fermion cloud size is about 80 μm×170 μm. c) Schematic drawing of experimental setup where a laser pulse with polarization perpendicular to the long axis of the cloud illuminates the degenerate fermion gas along the cloud's short axis after a controlled adiabatic expansion. Because of adiabatic expansion, the temperature is significantly reduced and the momentum spreads in the *x*-, *y*-, and *z*-directions map onto the spatial distribution of the cloud. Along the short axis (*x* or *y* direction) the momentum spread is small, whereas in the *z*-direction the momentum spread is larger. Correspondingly, the coherence time $t_x$ and $t_y$ for the atomic polarization induced by the external pump field are longer than $t_z$.

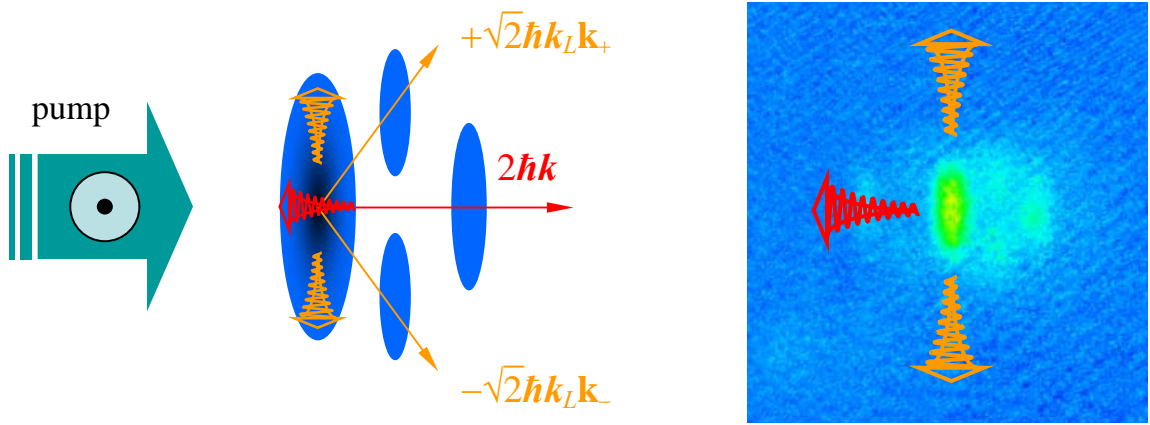

**Figure 2.** Demonstration of the co-existence of multiple mutual perpendicularly-propagating matter-optical wave-mixing processes in a momentum-squeezed, degenerate fermion gas. Here, the laser polarization is perpendicular to the long dimension of the cloud. In this case, longitudinal excitation (red wavy arrow) and transverse excitation (orange wavy arrows) processes co-exist. The longitudinal excitation leads to a forward collective atomic recoil mode, whereas the co-existing transverse excitation produces the usual $\pm 45°$ scattering components. The faint backscattered spot is attributed to a weak reflection of the pump light by the wall of the chamber. For this image the laser detuning is $\delta/2\pi = -358$ MHz, pump power = 5 mW, beam waist = 1 mm, TOF=15 ms, and the pump-pulse duration is 150 μs.

$\delta/2\pi=+500$MHz, 0.025 ms

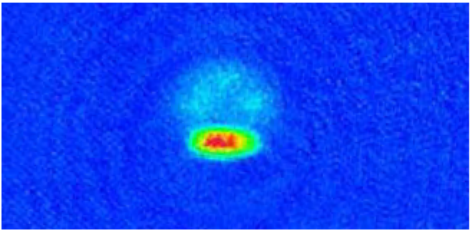

$\delta/2\pi=+500$MHz, 0.1 ms

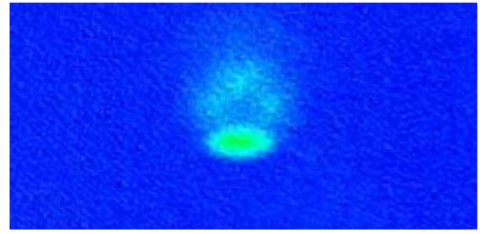

$\delta/2\pi=-500$MHz, 0.05 ms

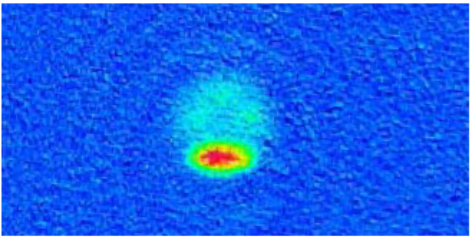

$\delta/2\pi=-500$MHz, 0.1 ms

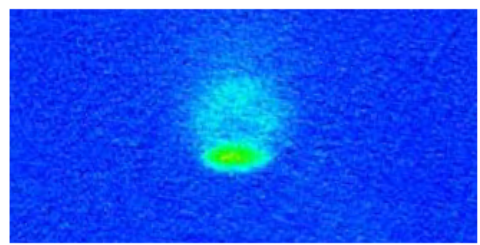

**Figure 3.** TOF images of collective atomic recoil motion modes with blue-(top) and red-detuned (bottom) pump lasers for different pulse durations. For all images the pump laser power is 2 mW, detuning is $|\delta/2\pi|=500$ MHz with respect to the $F'=9/2$ state in the $4P_{3/2}$ manifold, pulse duration is given in each image, TOF is 15 ms, and image size is about 70 μm×150 μm, and the apparatus was improved to eliminate any back reflections of the pump. The images for the red and blue detunings are nearly identical, indicating red-blue detuning symmetry in the generation of the collective recoil motion modes. Note that in the case of BECs, this symmetry has been shown to be broken.